\documentstyle[twocolumn,aps,epsf,prb]{revtex}
\begin{document}
\draft

\twocolumn[
\hsize\textwidth\columnwidth\hsize\csname@twocolumnfalse\endcsname

\title{Macroscopic resonant tunneling of magnetic flux}  

\author{D.V. Averin, Jonathan R. Friedman, and J.E. Lukens} 

\address{Department of Physics and Astronomy, SUNY at Stony Brook, 
Stony Brook NY 11794}

\date{\today}
\maketitle
\begin{abstract}

We have developed a quantitative theory of resonant tunneling of 
magnetic flux between discrete macroscopically distinct quantum 
states in SQUID systems. The theory is based on the standard 
density-matrix approach. Its new elements include the discussion 
of the two different relaxation mechanisms that exist for the 
double-well potential, and description of the ``photon-assisted'' 
tunneling driven by external rf radiation. It is shown that in 
the case of coherent flux dynamics, rf radiation should lead to 
splitting of the peaks of resonant flux tunneling, indicating 
that the resonant tunneling is a convenient tool for studying 
macroscopic quantum coherence of flux.   

\end{abstract}

\vspace*{3ex}

]

\section{Introduction}  \label{sec:intr} 

\vspace{1ex} 

Bose condensates of Cooper pairs in superconductors have a  
remakable ability to populate a single quantum state with a 
macroscopically large number of particles. This property of the 
Cooper pair condensates leads to macroscopic quantum effects in 
the dynamics of Josephson junctions, i.e. makes it possible for 
a Josephson junction to behave as a pure quantum state of a 
simple quantum mechanical system while still containing 
macroscopically large number of Cooper pairs. Due to the strongly 
nonlinear character of the Josephson dynamics even in 
the classical regime, macroscopic quantum effects can be quite 
non-trivial. They have been known for the past twenty years 
-- see reviews in \cite{mqt}, and are continuing to attract 
considerable interest \cite{nap}. This interest is to a large 
extent stimulated by the challenge to our understanding of the 
foundations of the quantum theory presented by direct 
manifestations of quantum mechanics at the macroscopic level 
\cite{qm1,qm2,qm3}. Another recent motivation for studying 
quantum effect in Josephson dynamics is provided by possible  
applications to quantum computation \cite{qc1,qc2,qc3,qc4}.   

The most advanced macroscopic quantum effect observed 
experimentally up to now is resonant tunneling between quantized 
energy levels in the adjacent wells of the Josephson potential 
\cite{rt1,rt2}. The aim of this work is to develop a theory 
of this phenomenon in SQUID systems, where the potential contains 
two wells each with a different value of the average 
magnetic flux. We consider the regime of weak energy dissipation 
important for studying coherent effects in resonant tunneling. 
This regime has not been discussed appropriately in the 
existing treatments of resonant tunneling in double-well 
potentials \cite{sq1,sq2,sq3} or multi-well potentials 
corresponding to the current-biased Josephson junctions \cite{c1,c2}.   
The most essential new feature of our approach is an account of 
the two types of relaxation mechanisms, intrawell and 
interwell, that exist in the system. The two 
relaxation mechanisms are very different in their dependence on 
the parameters of the SQUID potential, and lead to different   
shapes of the resonant tunneling peaks. Differences in 
relaxation mechanisms also make macroscopic resonant tunneling 
of flux different from the otherwise very similar ``mesoscopic'' 
resonant tunneling between charge states of small Josephson 
junctions \cite{cp1,cp2} and electron states in quantum dots 
\cite{qd1,qd2,qd3}. 

Another new element of this work is the discussion of 
the ``photon-assisted'' macroscopic resonant flux tunneling under 
rf irradiation. We show that in contrast to tunneling under 
stationary-bias conditions, the peaks of the photon-assisted 
tunneling depend qualitatively on the strength of decoherence 
in the flux dynamics. In the case of coherent flux dynamics, 
the resonant peaks of the photon-assisted tunneling are split 
in two. The splitting reflects the coherent hybridization of 
the macroscopic flux states in the two wells of the SQUID 
potential and is suppressed with increasing rate of decoherence. 
Very recently, such a splitting of the resonant flux-tunneling 
peaks has been observed experimentally \cite{sb}, demonstrating 
the quantum coherence of the macroscopically distinct flux 
states\cite{sb,de}. 
The paper is organized as follows. In Sec.\ 2 we derive the 
evolution equations for the density matrix describing the 
resonant flux tunneling under stationary-bias conditions, 
and introduce the two relaxation mechanisms for tunneling 
dynamics. Using these equations, we calculate the rate of 
flux tunneling in Sec.\ 3. In Sec.\ 4, we extend the results 
of Secs.\ 2 and 3 to the case of photon-assisted tunneling.

\section{Equations for the density matrix}  
\label{sec:mod} 

\vspace{1ex} 

To derive equations for the density matrix in the regime of 
resonant tunneling of magnetic flux $\Phi$ we consider the 
standard model 
of the phase dynamics in SQUIDS. The combination of the magnetic 
energy of the SQUID loop biased with an external flux and the 
Josephson coupling energy of the SQUID junctions (for details, see 
e.g., \cite{lik}) produces the double-well potential $U(\Phi)$ for 
$\Phi$ evolution (shown schematically in Fig.\ 1 below). The main 
part of the Hamiltonian governing the flux dynamics consists of the 
potential $U(\Phi)$ and the charging energy of the junction 
capacitance $C$: 
\begin{equation} 
H_0= \frac{Q^2}{2C} + U(\Phi) \, . 
\label{1} \end{equation} 
The charge $Q$ on the junction capacitance and the flux $\Phi$ 
satisfy standard commutation relations $[\Phi,Q]=i\hbar$. 
  
The two wells of the potential $U(\Phi)$ have discrete energy 
states $\varepsilon_{jn}$ with characteristic energy separation 
on the order of $\omega_j$, where $j=1,2$ is the well index,  
$\omega_j$ are the oscillation frequencies around the potential 
minima, and $n=0,1,...$ numbers the states within each well. 
The two frequencies $\omega_j$ have the same order of magnitude  
$\omega_1 \sim \omega_2 \equiv \omega_p$. External 
magnetic flux controls the energy difference between the states 
in opposite wells. Away from the resonance conditions, when all  
the energies $\varepsilon_{jn}$ are separated by large energy 
gaps of order $\omega_p$, the states $|jn\rangle$ are  
localized within the $j$th well, and the amplitude of the 
wavefunctions $\psi_{jn}(\Phi)$ in the opposite well is very 
small. However, when the energies of the two states $|1\rangle 
\equiv |1n_1\rangle$ and $|2\rangle \equiv |2n_2\rangle$ are close, 
$|\varepsilon | \ll \omega_p$, where $\varepsilon \equiv 
\varepsilon_{1n_1}- \varepsilon_{2n_2}$, these states become 
strongly coupled, and the wavefunctions spread over the both 
wells. As shown in the Appendix, strong coupling of the states 
$|1,2\rangle$ at resonance can be described by the tunneling 
amplitude $\Delta$, and the Hamiltonian (\ref{1}) reduces to the 
regular two-state form in the basis of these states: 
\begin{equation} 
H_0= \frac{1}{2} [\varepsilon (|1\rangle \langle 1| - 
|2\rangle \langle 2| ) - \Delta ( |1\rangle \langle 2| + 
|2\rangle \langle 1|)] \, ,   
\label{2} \end{equation} 

\vspace*{-4ex} 

\[ \Delta= (\omega_1 \omega_2)^{1/2}D/\pi \, , \] 
where $D$ is the quantum mechanical transparency of the barrier 
separating the wells.  

Without perturbations, the two-state dynamics described by (\ref{2}) 
is decoupled from the other states of the Hamiltonian 
(\ref{1}). The most important perturbation creating such 
coupling is the energy dissipation that induces transitions 
between the states $|1,2\rangle$ and other states $|jn\rangle$. 
In the relevant temperature range below superconducting energy 
gap of the junction electrodes, the quasiparticle tunneling is 
suppressed and the main source of energy dissipation is the 
electromagnetic environment of the system. Under the  
assumption that the electromagnetic modes of the environment are 
in equilibrium at temperature $T$, and are well described by 
linear electrodynamics, the interaction between the flux $\Phi$ 
and the heat bath of these modes can be written as 
\begin{equation}
V=-I_f \Phi \, .  
\label{3} \end{equation} 
Here $I_f$ is the fluctuating current created in the SQUID 
loop by the environment, with the correlation function given 
by the fluctuation-dissipation theorem: 
\begin{equation} 
\langle I_f(t) I_f(t+\tau) \rangle = \int \frac{d \omega }{\pi} 
\frac{\omega G(\omega) e^{i\omega \tau } }{1-e^{- \omega /T} } \, , 
\label{4} \end{equation}
where the brackets $\langle ... \rangle$ denote averaging over 
the equilibrium density matrix of the environment, and $G(\omega)$ 
is the dissipative part of the environment 
conductance. Equation (\ref{4}) is sufficient to characterize 
completely the effects of the weak energy dissipation considered 
in this work. For arbitrary dissipation strength, one can use 
the Caldeira-Leggett model \cite{cl} to express explicitly the 
environment Hamiltonian and the current operator $I_f$ in terms 
of a set of harmonic oscillators. 

The interaction (\ref{3}) induces both ``vertical'' transitions 
within each well and direct interwell transitions. In terms of the 
two-state dynamics with the Hamiltonian (\ref{2}), the latter 
correspond to the modulation of the tunneling amplitude $\Delta$ 
by the environment. The matrix elements of this type of interwell 
transitions are, however, smaller by a factor of $\Delta/\omega_p$ 
than those of the intrawell transitions. The small matrix 
elements can be neglected under the conditions of resonance, when 
the flux tunneling between the two wells is dominated by the 
stronger resonant processes. In this approximation, we can omit  
the terms in the flux operator in the interaction (\ref{3}) that 
are non-diagonal in the well index $j$: 
\begin{equation}  
\Phi =\sum_{j,n,n'} \Phi^{(j)}_{n,n'}|jn\rangle \langle jn'| 
\, . 
\label{5} \end{equation}
Here $\Phi^{(j)}_{n,n'}$ are the matrix elements of $\Phi$ in 
the $|jn\rangle$ basis. 
The perturbation (\ref{3}) with the flux operator (\ref{5}) has 
two effects on the dynamics of the states $|1,2\rangle$. The first is 
fluctuations of the energy difference $\varepsilon$, which induce 
transitions between these states and lead to the loss of mutual 
coherence between them. The part of the $\Phi$ operator (\ref{5}) 
responsible for these fluctuations can be written as $\delta \Phi 
(|1\rangle \langle 1| - |2\rangle \langle 2|)$, where $\delta \Phi$ 
is half of the difference between the average flux values in the 
states $|1\rangle$ and $|2\rangle$. The remaining terms in (\ref{5}) 
induce intrawell transitions from the states $|1\rangle$ and 
$|2\rangle$ to the other states in the corresponding wells. 

For weak dissipation both effects can be described quantitatively 
by the standard density matrix technique -- see, e.g., \cite{bl}. 
The description starts from the equation for the evolution of the 
density matrix $\rho$, obtained treating the coupling $V$ (\ref{3}) 
in second-order perturbation theory: 
\begin{equation} 
\dot{\rho } (t) = - i [H_0,\rho] - \int^{t} d\tau \langle 
[V(t), [V(\tau), \rho(\tau)]] \rangle \, . 
\label{6} \end{equation}
If the environment has a large cut-off frequency $\omega_c \gg 
\varepsilon, \Delta$, the density 
matrix evolves slowly on the time scale of variations of 
$V(t)$, and we can make the Markov approximation $\rho (\tau ) 
\simeq \rho (t)$ in the last term of eq.\ (\ref{6}). 
The condition of weak dissipation also allows us to keep only 
the dissipative terms in eq.\ (\ref{6}) that do not oscillate 
in time with frequencies of the main Hamiltonian $H_0$, since 
only these terms lead to effects that accumulate with time. 
Using these approximations to evaluate the dissipative part 
of the eq.\ (\ref{6}), we obtain the final equation for the 
density matrix in the basis of resonant states $|1,2\rangle$ 
relevant for the transfer of flux between the wells: 
\begin{equation}
\dot{\rho} = - i [H_0,\rho]+ \Gamma[\rho] + 
\gamma[\rho] \, .  
\label{7} \end{equation}
The term $\Gamma$ in this equation describes the effect of 
the intrawell transitions from the states $|1\rangle$ 
and $|2\rangle$: 
\begin{equation}
\Gamma[\rho]= -\left( \begin{array}{cc}
\Gamma_1 \rho_{11} \, , & (\Gamma_1+\Gamma_2)\rho_{12}/2  \\
(\Gamma_1+\Gamma_2)\rho_{21} /2 \, , & 
\Gamma_2 \rho_{22} \end{array} \right) \, .
\label{8} \end{equation}
At temperatures smaller than energy separation in the wells, 
the total decay rate $\Gamma_j$ in eq.\ (\ref{8}) of the state 
$|j\rangle $ into the states with lower energy in the same well is: 
\[ \Gamma_j = \sum_{n<n_j} \frac{2 |\Phi^{(j)}_{n,n_j}|^2 
}{\hbar^2} (\varepsilon_{n_j}- \varepsilon_n) G(\varepsilon_{n_j}- 
\varepsilon_n) \, . \]

The second dissipation term $\gamma [\rho]$ in eq.\ (\ref{7}) 
describes transitions and decoherence within the $|1\rangle$, 
$|2\rangle$ subspace, and has a simple form in the basis 
of energy eigenstates of the two-state Hamiltonian (\ref{2}):  
\begin{equation} 
\gamma [\rho]= - U^{\dagger} \left( \begin{array}{cc}
\gamma_u r_{11}- \gamma_d r_{22}  \, , & (\gamma+ \frac{\gamma_d+
\gamma_u}{2} )r_{12}  \\ (\gamma+ \frac{\gamma_d+\gamma_u}{2}) 
r_{21} \, ,& \gamma_d r_{22}-\gamma_u r_{11} \end{array} \right) 
U \, . 
\label{9} \end{equation}
Here $r$ is the density matrix in the eigenstate basis: $R=U 
\rho U^{\dagger}$, and $U$ is the rotation matrix from this basis 
to the flux basis $|1\rangle$, $|2\rangle$: 
\[ U= [(1- \varepsilon /\Omega)^{1/2} \sigma_z + (1+\varepsilon / 
\Omega)^{1/2} \sigma_x]/\sqrt{2} \, ,\] 
where the $\sigma$'s denote Pauli matrices, and $\Omega \equiv 
(\varepsilon^2+ \Delta^2 )^{1/2} $. The transition rates 
$\gamma_{d,u}$ and the decoherence rate $\gamma$ are: 
\[ \gamma_d =\frac{g \Delta^2}{\Omega } \frac{1}{1-e^{-\Omega /T}} 
\,  \;\;\; \gamma_u = \gamma_d e^{-\Omega /T} \, , \;\;\; 
\gamma = 2g T \frac{\varepsilon^2}{\Omega^2} \, ,  \]   
with the dimensionless parameter $g=2G(\delta \Phi)^2/\hbar$
characterizing the strength of the interwell relaxation, and 
we assumed that $G(\omega)$ is constant in the small-frequency 
range $\omega \sim \Omega$. 

Equation (\ref{7}), with the dissipation terms (\ref{8}) and 
(\ref{9}), is used below to describe resonant tunneling of flux 
between the two wells in various regimes. Before doing this, we 
discuss the relative magnitude of the two 
dissipation terms in this equation. Since the width of the wells 
is of the same order of magnitude as the barrier between them, 
the magnitude of the flux matrix elements of the intrawell 
relaxation $\Gamma[\rho]$, and of the interwell relaxation 
$\gamma[\rho]$ (determined, respectively, by the ``width'' of 
the wavefunctions inside the wells and the distance between 
the wells) should be close. 
The main difference between the two relaxation mechanisms  
is that the intrawell transitions dissipate energy $\omega_j$, 
whereas the interwell relaxation $\gamma[\rho]$ involves only 
much smaller energies on the order of $\varepsilon, \,\Delta, 
\,T$. This means that the intrawell relaxation typically 
dominates the flux-tunneling dynamics. In 
particular, even under the assumed condition of weak relaxation  
(which for $\Gamma[\rho]$ means that the rates $\Gamma_j$ are 
small compared to the oscillation frequencies $\omega_j$) the 
rates $\Gamma_j$ can still be much larger than the frequencies 
$\varepsilon, \, \Delta$ of the two-state dynamics. 
The interwell relaxation $\gamma[\rho]$ will generally only be 
stronger than the intrawell relaxation if the environment has 
relatively low cut-off frequency $\omega_c \ll \omega_p$.

\section{Stationary bias}  
\label{sec:stat} 

\vspace{1ex} 

In this Section, we calculate the rate of the resonant tunneling 
between the wells in the situation when the external flux 
through the SQUID loop does not contain an ac component, and the  
energy dissipation drives the initial flux state in the left well 
towards equilibrium. At low temperatures $T\ll 
\omega_p$ the flux stays in the ground state of the left well 
and tunnels into the right well out of this state (Fig.\ 1). In 
this case, the relaxation rate $\Gamma_1$ obviously vanishes. 
We begin by considering the situation with only the intrawell 
relaxation present. Equation (\ref{7}) can then be 
written in matrix elements as:    
\begin{eqnarray}   
\dot{\rho}_{11} & = &  \Delta \mbox{Im} \rho_{12} 
\, , \;\;\;\; \dot{\rho}_{22}  =  -\Delta \mbox{Im} \rho_{12} 
-2 \Gamma \rho_{22} \, ,  \nonumber \\  
\dot{\rho}_{12} & = & - (i \varepsilon + \Gamma ) 
\rho_{12} + (i \Delta/2)(\rho_{22}-\rho_{11}) \, , \label{10}    
\end{eqnarray}  
with $\Gamma_2/2\equiv \Gamma$ in this Section.

\begin{figure}[htb]
\setlength{\unitlength}{1.0in}
\begin{picture}(3.,2.3) 
\put(.0,.0){\epsfxsize=3.in\epsfysize=2.3in\epsfbox{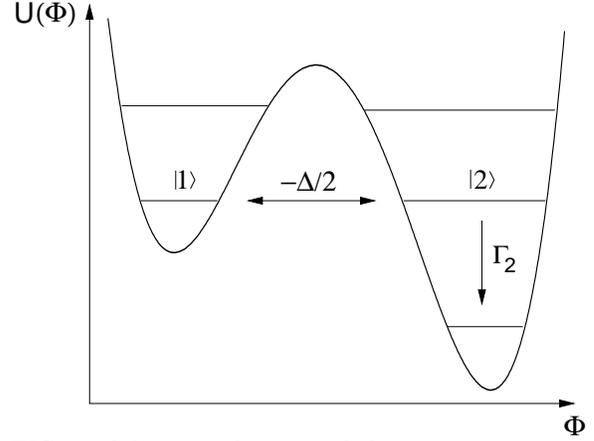}}
\end{picture}

\caption{Schematic diagram of the macroscopic resonant tunneling 
of flux $\Phi$ in the double-well potential $U(\Phi)$ at low 
temperatures under stationary-bias conditions. The flux 
tunnels out of ground state $|1\rangle$ in the left well coupled 
with the amplitude $-\Delta/2$ to the resonant state $|2\rangle$ 
in the right well, where it decays with the rate $\Gamma_2$ into 
the lower states in this well. }

\end{figure} 

After transformation to the real and imaginary parts of the 
off-diagonal matrix element $\rho_{12}$ and the sum/difference 
$\rho_{11} \pm \rho_{22}$ of the diagonal elements of the density 
matrix, eqs.\ (\ref{10}) can be solved directly with the initial 
condition that the flux is in the left well at time $t=0$, 
$\rho_{11}(0)=1$: 
\begin{eqnarray}  
\rho_{11}(t) =  \frac{1}{2} \frac{e^{-\Gamma t}}{\omega^2 + 
\lambda^2} \left[ (\Gamma^2 + \omega^2) ((1+ \frac{\lambda^2}{ 
\Gamma^2} ) \cosh \lambda t + \right. \nonumber \\  
\left. 2 \frac{\lambda}{\Gamma} \sinh 
\lambda t) + (\lambda^2 -\Gamma^2) ((1-\frac{\omega^2}{ \Gamma^2}) 
 \cos \omega t -2 \frac{\omega}{ \Gamma} \sin \omega t ) 
\right] \, , \nonumber \\ 
\rho_{22}(t) = \frac{1}{2} \frac{e^{-\Gamma t}}{\omega^2 + 
\lambda^2} \left[ (\Gamma^2 + \omega^2) (1- \frac{\lambda^2}{\Gamma^2} 
) \cosh \lambda t +  \right. \nonumber \\ 
 \left. (\lambda^2 -\Gamma^2) (1+ \frac{\omega^2}{ \Gamma^2}) 
\cos \omega t \right] \, . \label{11} 
\end{eqnarray} 
In eq.\ (\ref{11}), the eigenfrequencies $\omega$ and $\lambda$ of 
the system of equations (\ref{10}) are: 
\begin{equation} 
\omega \, , \lambda =  \left[ ( \frac{(\Omega^2 - 
\Gamma^2)^2}{4} + \Gamma^2 \varepsilon^2 )^{1/2} \pm  
\frac{\Omega^2 - \Gamma^2}{2} \right]^{1/2}  \, . 
\label{12} \end{equation} 
Equations (\ref{11}) contain all the information about dynamics 
of the flux tunneling. When the relaxation rate $\Gamma_2$ is much 
smaller than the oscillation frequency $\Omega$, the 
tunneling process consists of the weakly damped coherent 
oscillations of the flux between the wells followed by relaxation 
in the right well. With increasing relaxation rate the oscillation 
part of this process becomes increasingly more damped and turns 
into incoherent jumps of the flux from the left into the right 
well represented by the non-oscillatory exponential decay of 
$\rho_{11}$.  

The matrix element $\rho_{11}(t)$ has the meaning of the 
probability for the flux to remain in the left well at time $t$.
In the case of monotonous decay of $\rho_{11}$, its derivative  
$f(t)= -[\rho_{11}(t)]'$ gives the probability density of the 
time for flux tunneling between the wells. This probability 
density can be used to find all statistical characteristics of 
the flux tunneling process. For instance, the average time $\tau$ 
it takes for the flux to tunnel can be calculated as $\tau = 
\int_0^{\infty} dt~ t f(t) =\int_0^{\infty} dt \rho_{11}(t)$. 
In the coherent regime, when $\rho_{11}(t)$ oscillates in time, 
$f(t)$ can be negative and cannot be interpreted as probability 
density, and the question of how to define the tunneling time $\tau$  
becomes non-trivial. To define $\tau$, one needs to establish the 
event that terminates the tunneling process. The definition adopted 
below assumes that the tunneling process ends when the flux makes 
the transition from the state $|2\rangle$ into one of the lower 
energy states in the right well. This definition is motivated by 
the fact that such a transition eliminates the possibility for the 
flux to return into the left well. With such definition, the time 
the flux spends in the state $|2\rangle$ is included in the 
tunneling time $\tau$ which then should be calculated as 
\begin{equation} 
\tau =\int_0^{\infty} dt (\rho_{11}(t)+\rho_{22}(t)) \, . 
\label{13} \end{equation}

From equations (\ref{11}) and (\ref{13}) we obtain the tunneling 
rate: 
\begin{equation}  
\tau^{-1} = \frac{\Delta^2 \Gamma_2 }{2\Delta^2 + \Gamma_2^2+ 
4 \varepsilon^2 }\, .  
\label{14} \end{equation}  
Equation (\ref{14}) describes the Lorentzian peak of the resonant 
flux tunneling. It shows that the resonant flux tunneling under 
stationary-bias conditions does not allow one to distinguish 
qualitatively between the regimes of coherent and incoherent flux 
tunneling since the shape of the resonance peak (\ref{14}) remains 
the same regardless of the magnitude of the relaxation/decoherence 
rate $\Gamma$.

The average tunneling rate $\tau^{-1}$ can be calculated without 
explicit solution of the time-dependent equations for the density 
matrix. Instead of attempting to describe an individual tunneling 
event with the time-dependent solution, we can consider a large 
number of these events, assuming that after each transition from the 
left to the right well the system is immediately returned back to 
its initial state and the process is repeated. This immediate return 
means that the system is effectively decaying from the resonant 
state in the right well directly into the initial state in the left 
well, and can be modeled by adding the term $\Gamma_2 \rho_{22}$ 
into the equation for $\rho_{11}$. With such a modification, eq.\ 
(\ref{7}) has a non-trivial stationary solution $\rho^{(0)}$, and 
the tunneling rate $\tau^{-1}$ defined by eq.\ (\ref{13}) can be 
found from this solution as 
\begin{equation} 
\tau^{-1}= \Gamma_2 \rho_{22}^{(0)}\, .    
\label{tau} \end{equation}  
Since this method does not require solution of the time-dependent 
equations for the density matrix, it considerably simplifies the 
calculation of the average tunneling rate.    
 
To illustrate this procedure, we consider first the same tunneling 
under the stationary bias conditions described by eqs.\ (\ref{10}) 
with the term $\Gamma_2 \rho_{22}$ included into the equation for 
$\rho_{11}$. Solving the stationary equation for the off-diagonal 
element $\rho_{12}$ of the density matrix and plugging the solution 
into the equations for the diagonal elements, we get the simple rate 
equations:   
\begin{equation} 
\dot{\rho}_{11} = \Gamma' (\rho_{22}-\rho_{11}) + \Gamma_2 
\rho_{22} \, ,\;\;\;\; \dot{\rho}_{22} = - \dot{\rho}_{11}\, , 
\label{16} \end{equation} 
where the transfer rate 
$\Gamma' =\Delta^2 \Gamma_2 /(\Gamma_2^2+ 4 \varepsilon^2)$    
between the two wells can be viewed as the ``Golden-rule'' rate of 
transition with the matrix element $\Delta /2$ into the state 
$|2\rangle$ broadened by the relaxation $\Gamma_2$. From the 
stationary solution of eq.\ (\ref{16}) we find that $\rho_{22}^{(0)} 
= \Gamma' /(\Gamma_2+2\Gamma')$, and see that eq.\ 
(\ref{tau}) indeed reproduces the tunneling rate (\ref{14}).  
 
The tails of the resonant peak (\ref{14}) at $\varepsilon \gg 
\Delta, \Gamma_2$ allow for another simple interpretation. 
At large $\varepsilon$, the wavefunction of the eigenstate of 
the two-state Hamiltonian localized in the left well has the 
probability amplitude $\Delta/2\varepsilon $ in the right well. The 
tunneling rate $\bar{\Gamma}$ in this regime can be found then as 
the probability to be in the right well times the relaxation rate 
$\Gamma_2$: 
\begin{equation} 
\bar{\Gamma}=\Gamma_2 \Delta^2 /4\varepsilon^2\, . 
\label{16a} \end{equation} 
This simple reasoning indeed reproduces the tails of the peak 
(\ref{14}) and allows us to 
obtain an estimate of the tunneling rate between the resonances. 
As shown in the Appendix, the wavefunction amplitude in the 
right well between the resonances is $\pi \Delta/[2\omega_2 
\sin (\pi \varepsilon/\omega_2)]$. From this we can write: 
\begin{equation}
\tau^{-1} = \Gamma_2 \left(\frac{\pi \Delta }{2\omega_2  
\sin (\pi \varepsilon/\omega_2) } \right)^2 \, .  
\label{17} \end{equation} 
It should be noted that eq.\ (\ref{17}) is only an estimate, since 
$\Gamma_2$ and $\Delta$ depend on $\varepsilon$ for $\varepsilon 
\sim \omega_2$, and can be different from their values at resonance. 
However, at $\varepsilon \ll \omega_2$, they are constant, and eqs.\ 
(\ref{14}) and (\ref{17}) coincide for $\varepsilon \gg \Delta, 
\Gamma_2$. In this range of $\varepsilon$ the tunneling rate 
$\tau^{-1}$ changes as $\varepsilon^{-2}$. 

If the interwell relaxation is non-negligible, we need to keep 
both relaxation terms in eq.\ (\ref{7}). The assumption 
that the relaxation rates are small in comparison with $\Delta$, 
allows us to use eq.\ (\ref{9}) for the interwell relaxation and 
makes it convenient to consider the flux dynamics in the 
eigenstates basis. The stationary values of the off-diagonal 
elements of the density matrix $r$ in this basis are vanishing 
for weak relaxation. To find the diagonal elements of $r$, we 
transform eq.\ (\ref{8}) for the intrawell relaxation (with 
$\Gamma_1=0$ and added term $\Gamma_2 \rho_{22}$ in the evolution 
of $\rho_{11}$) into this basis. Neglecting rapidly oscillating 
terms, we see that the diagonal part of the weak intrawell 
relaxation is: 
\[ \dot{r}_{11} = \Gamma_2 [ \frac{-\varepsilon}{2\Omega} + 
\frac{1}{4} (1+\frac{\varepsilon^2}{\Omega^2}) (r_{22}-r_{11}) ] 
 \, , \;\;\;\; \dot{r}_{22}=-\dot{r}_{11} \, .\] 
Combining this expression with eq.\ (\ref{9}) for the interwell 
relaxation, we find the stationary values of the diagonal elements 
of $r$. After transformation back to the flux basis we finally 
obtain the stationary element $\rho_{22}^{(0)}$ of the density 
matrix $\rho$ in the flux basis and the flux tunneling rate 
(\ref{tau}): 
\begin{equation} 
\tau^{-1}= \frac{\Gamma_2}{2} 
\frac{\Omega \coth (\Omega/2T) +\varepsilon +\mu}{\Omega 
\coth (\Omega/2T) +\mu (1+2 \varepsilon^2 /\Delta^2 )} \, , 
\label{18} \end{equation}  

\vspace{-2ex}

\[ \mu \equiv \Gamma_2/2g \, . \]

The parameter $\mu$ in eq. (\ref{18}) can be interpreted as the 
energy at which the characteristic interwell relaxation rate (which 
increases with the energy difference between the two resonant 
states) becomes equal to the rate 
$\Gamma_2$ of the relaxation in the right well. The tunneling 
rate (\ref{18}) is plotted in Fig.\ 2 for zero temperature 
and several values of $\mu$. The interwell relaxation becomes 
stronger with decreasing $\mu$, making the the resonant-tunneling 
peak progressively more asymmetric. For negative bias 
$\varepsilon$, when the state $|2\rangle$ in the right well 
is higher in energy than the state $|1\rangle$ in the left well, 
the interwell relaxation is suppressed for low temperatures, $T\ll 
\Delta$, and  the tail of the resonant peak (\ref{18}) coincides 
with that of the Lorentzian peak (\ref{14}). On the other hand, 
for positive $\varepsilon$, transitions from the state $|1\rangle$ 
into $|2\rangle$ are allowed, and if $\mu \leq \Delta$, the 
interwell relaxation dominates for all positive $\varepsilon$. 
The tunneling rate decreases in this case only as $1/\varepsilon$ 
with increasing $\varepsilon$. 

At $\varepsilon \gg \Delta$, and vanishing temperature $T$, the 
tunneling rate (\ref{18}) determined by the interplay between the 
interwell and intrawell relaxation can be understood in terms of the 
competition between the two tunneling paths (see inset in Fig.\ 2). 
One is the direct decay within the right well out of the eigenstate 
of the two-state Hamiltonian localized predominantly 
in the left well, but with a small probability amplitude in the 
right well. The rate of this decay is $\bar{\Gamma}$ (\ref{16a}). 
Another is the transition between the two eigenstates induced by 
the interwell relaxation that transfers the probability between 
the two wells and is followed by the intrawell decay out of the 
lower-energy eigenstate with the rate $\Gamma_2$. The rate of the 
interwell transition between the eigenstates is 
\begin{equation} 
\bar{\gamma}= \frac{g \Delta^2 }{\varepsilon} \, .
\label{19} \end{equation} 
At $\mu < \Delta$, the second path dominates, and for sufficiently 
large $\varepsilon$, $\varepsilon \gg \Delta^2/\mu$, the bottleneck 
of the relaxation process is the interwell transition between the 
eigenstates and the tunneling rate (\ref{18}) becomes independent of 
the intrawell rate $\Gamma_2$: $\tau^{-1}= \bar{\gamma}$.  
In general, the competition between the two types of transitions 
gives an expression for the tunneling rate, $\tau^{-1}= \Gamma_2 
(\bar{\gamma}+ \bar{\Gamma})/(\bar{\gamma}+\Gamma_2)$, that agrees 
with eq.\ (\ref{18}) at $\varepsilon \gg \Delta$. 

For large $\mu$, $\mu \gg \Delta$, the interwell relaxation is weak 
close to resonance, and for sufficiently small bias, $\varepsilon 
\ll \mu$, eq.\ (\ref{18}) coincides with eq.\ (\ref{14}) with  
$\Gamma_2 \ll \Delta$. However, at larger bias, $\varepsilon \gg \mu$, 
the interwell relaxation increases and the tunneling rate is again 
given by $\bar{\gamma}$ (\ref{19}). Only when $\mu$ becomes comparable 
to the level separation $\omega_p$ in the wells, the interwell 
relaxation is completely negligible and the resonant tunneling peak has 
the Lorentzian shape for all relevant energies.

\begin{figure}[htb]
\setlength{\unitlength}{1.0in}
\begin{picture}(3.0,2.3) 
\put(0.,0.){\epsfxsize=3.0in\epsfysize=2.3in\epsfbox{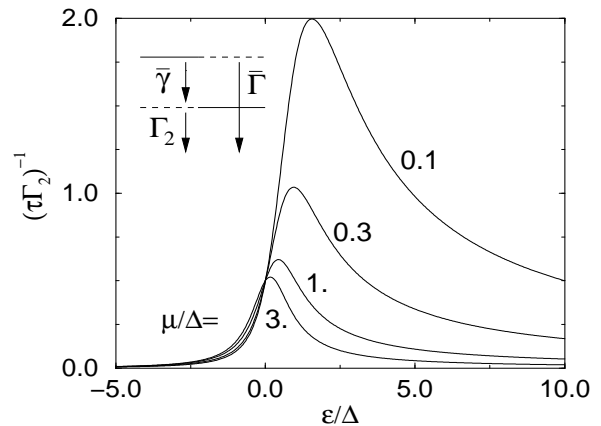}}
\end{picture}
\caption{The zero-temperature rate $\tau^{-1}$ of flux tunneling 
between the wells of the double-well potential as a function of the 
energy bias $\varepsilon$ for different values of $\mu$, which 
characterizes the relative strength of the two relaxation mechanisms 
\protect (\ref{18}). The inset shows the diagram of the off-resonant 
transitions. }
\end{figure}

Away from resonance, when $\varepsilon \sim \omega_p$, the 
interwell tunneling with the rate (\ref{19}) 
corresponds to the transitions between the states localized 
in opposite wells that are the closest in energy, while ``intrawell'' 
tunneling (\ref{17}) corresponds to the transition between the 
states that are at least next-nearest neighbors in energy. Although 
there is no qualitative difference between the two types of the 
transitions away from the resonance, they lead to very different 
shapes for the tunneling peaks close to resonance.

\section{ Photon-assisted tunneling}  
\label{sec:rf} 

When the SQUID is irradiated with an external rf signal, the 
macroscopic resonant flux tunneling can go more effectively 
through one of the excited states in the left well of the SQUID 
potential rather than out of the ground state, since the amplitude 
of tunneling $-\Delta/2$ out of the excited state is much larger 
than the tunneling amplitude for the ground state. In this 
Section, we consider the situation when an rf signal of frequency 
$\omega$ resonantly couples the ground state $|0\rangle$ in the 
left well of the potential to an excited state $|1\rangle$ with 
energy $E$ in this well (Fig.\ 3). The energy $E$ is on the order 
of $\omega_p$, and the condition of the resonant excitation is 
that the detuning $\nu=E-\omega$ is small, $\nu \ll 
\omega_p$. If the amplitude $a$ of the rf excitation is also 
relatively small, $a\ll \omega_p$, the off-resonant coupling to 
other states\cite{off} is not important, and the coupling between 
the states $|0\rangle$ and $|1\rangle$ can be described in the 
rotating-wave approximation. In this approximation, the terms in the 
coupling that oscillate rapidly (with frequencies on the order of 
$\omega_p$) are neglected, and the coupling Hamiltonian is written 
as:   
\begin{equation}  
H_{rf} = \frac{a}{2} (|0\rangle \langle 1| e^{-i\nu t} + |1\rangle 
\langle 0| e^{+i\nu t}) \, . 
\label{r1} \end{equation} 
If the excited state $|1\rangle$ in the left well is coupled 
resonantly with amplitude $-\Delta/2$ to a state $|2\rangle$ in the 
right well that is shifted in energy by $\varepsilon$ with respect 
to $|1\rangle$, the total Hamiltonian for the flux dynamics in the 
basis of the three states $|0\rangle$, $|1\rangle$, and $|2\rangle$ 
is:  
\begin{equation} 
H_0=  \left( \begin{array}{ccc} 0 & a/2 & 0 \\ 
a/2 & \nu &  -\Delta/2\\ 0 & -\Delta/2 & \nu -\varepsilon 
 \end{array} \right) \, . 
\label{r2} \end{equation}  

\begin{figure}
\setlength{\unitlength}{1.0in}
\begin{picture}(3.,2.4) 
\put(.0,.0){\epsfxsize=3.in\epsfysize=2.4in\epsfbox{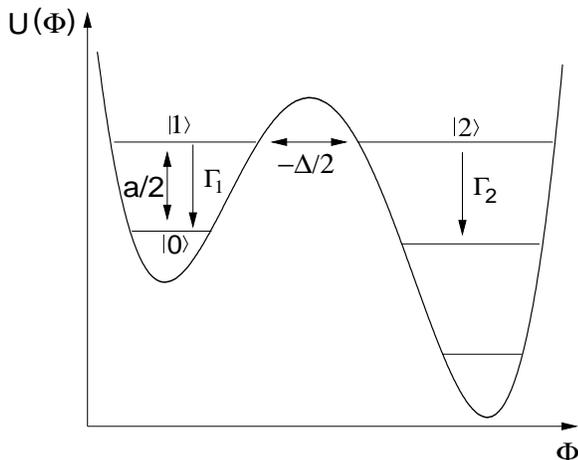}}
\end{picture}
\caption{Schematic diagram of the ``photon-assisted'' 
macroscopic resonant tunneling of flux stimulated by an rf 
perturbation of strength $a/2$. }  
\end{figure}

As in the previous Section, the average flux tunneling rate 
$\tau^{-1}$ can be calculated according to eq.\ (\ref{tau}) 
from the stationary density matrix $\rho$ of the system in the 
basis of states $|0\rangle$, $|1\rangle$, and $|2\rangle$. 
Time evolution of $\rho$ is described by the same eq.\ (\ref{7})  
but with the Hamiltonian $H_0$ given now by eq.\ (\ref{r2}).
We begin discussion of the dynamics of flux tunneling in the 
three-state system (\ref{r2}) with the case when the interwell 
relaxation $\gamma[\rho]$ can be neglected, and the only 
mechanism of the energy relaxation in the system is the intrawell 
relaxation $\Gamma[\rho]$. This relaxation is characterized by 
the two rates $\Gamma_{1,2}$ of the transitions from the states 
$|1,2\rangle$ into the lower energy states of the left and right 
potential well, respectively (Fig.\ 3). An obvious generalization 
of eq.\ (\ref{8}) for $\Gamma[\rho]$ to the three-state basis 
gives the off-diagonal part of eq.\ (\ref{7}) for the time 
evolution of $\rho$ of the following form: 
\begin{eqnarray} 
\dot{\rho}_{01} & = & (i \nu - \Gamma_1/2 )\rho_{01} +
ia(\rho_{00}-\rho_{11})/2 -i\Delta \rho_{02}/2  \, , 
\nonumber \\   
\dot{\rho}_{12} & = & -(i \varepsilon + (\Gamma_1+\Gamma_2)/2 ) 
\rho_{12} + i\Delta(\rho_{22}-\rho_{11})/2 -ia \rho_{02}/2  \, ,
\nonumber \\ 
\dot{\rho}_{02} & = & (i (\nu- \varepsilon) - \Gamma_2/2 ) 
\rho_{02} - ia\rho_{12}/2 -i\Delta \rho_{01}/2  \, . \label{r5}   
\end{eqnarray} 

Equations (\ref{r5}) allow us to express the off-diagonal elements of 
$\rho$ in terms of the diagonal ones in the stationary regime.  
Inserting the stationary values of the off-diagonal elements 
into the equations for the diagonal elements: 
\begin{eqnarray}
\dot{\rho}_{00} & = &  - a\mbox{Im} \rho_{01} + 
\Gamma_1\rho_{11} + \Gamma_2\rho_{22} \, ,\nonumber \\  
\dot{\rho}_{11} & = &  a\mbox{Im} \rho_{01} + \Delta 
\mbox{Im} \rho_{12} -\Gamma_1\rho_{11} \, , \label{r6} \\  
\dot{\rho}_{22} & = &  - \Delta \mbox{Im} \rho_{12}  - 
\Gamma_2\rho_{22} \, , \nonumber  
\end{eqnarray} 
we calculate the stationary probability $\rho_{22}^{(0)}$ and 
find the flux tunneling rate (\ref{tau}).  

Equations (\ref{r6}) were written under the assumption that the 
relaxation in the left well brings the system out of the state 
$|1\rangle$ directly into the ground state $|0\rangle$. Although
this is strictly true only in the case when $|1\rangle$ is the 
first excited state in the well, eqs.\ (\ref{r6}) can be also 
used to calculate the average flux tunneling rate in other 
situations. Indeed, when the rf signal drives the system into 
the state $|1\rangle$ that is not the first excited state, the 
intermediate states in the left well that exist between the 
states $|0\rangle$ and $|1\rangle$ are populated by the process 
of relaxation out of $|1\rangle$. If the flux tunneling out of 
these states is neglected (similarly to tunneling out of the 
state $|0\rangle$), their effect on the average tunneling rate 
can be accounted for by inclusion of the occupation 
probabilities of these states in the normalization condition. 
Since these probabilities in the stationary regime are 
proportional to the stationary occupation probability $\rho_{11}$, 
this can be done through an additional factor $\lambda$ in the 
normalization condition for the state $|0\rangle$, $|1\rangle$, 
and $|2\rangle$: 
\begin{equation} 
\rho_{00}+\lambda \rho_{11}+\rho_{22}=1 \, .
\label{rn} \end{equation} 
To give an example, we can calculate the factor $\lambda$ 
assuming that the left well is parabolic in the relevant energy 
range. Then, the standard result for the linear relaxation of the 
harmonic 
oscillator (see, e.g., \cite{zp}) is that the oscillator makes 
the transitions only between the nearest-neighbor states and 
that the transition rate from the state $|m\rangle$ into  
$|m-1\rangle$ is proportional to $m$. This means that in the 
stationary state $\rho_{m-1,m-1}=m\rho_{m,m}/(m-1)$, and if the 
rf radiation drives the system into the $n$th excited 
state of the left well, then $\lambda=n\sum_{m=1}^{n}(1/m)$. 
To avoid extra parameters, however, we assume from now on that 
the state $|1\rangle$ is the first excited state in the left 
well, so that $\lambda=1$. 

Sufficiently compact analytical expressions for the tunneling 
rate can be obtained from eqs.\ (\ref{r5}) and (\ref{r6}) only in 
certain limits. For example, for small rf amplitude and weak 
relaxation, $a\ll \Gamma_{1,2} \ll \Delta $, we get\cite{mes}: 
\begin{equation}
\tau^{-1} = \frac{\Gamma_2 a^2 \Delta^2}{
(2\nu-\varepsilon-\Omega)^2 (2\nu-\varepsilon+\Omega)^2 + 
4( \Gamma_1(\nu-\varepsilon)+\Gamma_2\nu)^2 } \, . 
\label{r7} \end{equation}
Equation (\ref{r7}) describes two peaks (discussed in more details 
below) in the dependence of the tunneling rate on the detuning 
$\nu$. The peaks are broadened by the relaxation, and their 
positions correspond to the two eigenstates of the Hamiltonian 
(\ref{2}): $\nu=(\varepsilon \pm \Omega) /2$. Another expression 
for the tunneling rate can be obtained for large relaxation rates 
$\Gamma_{1,2} \gg a,\Delta $:  
\begin{equation}
\tau^{-1} = \frac{\Gamma_2 a^2 \Delta^2}{
(4\nu^2+\Gamma_1^2) (4(\nu-\varepsilon)^2 +\Gamma_2^2) } \, . 
\label{r8} \end{equation}
Depending on the relation between the energy bias $\varepsilon$ 
and the relaxation rates, the tunneling rate (\ref{r8}) as a 
function of detuning $\nu$ contains either one (for small 
$\varepsilon$) or two separate peaks (for large $\varepsilon$). 
The peak positions in this case coincide with the position of 
the energy levels $|1,2\rangle$ localized in the two wells.

For arbitrary parameters it is convenient to use the stationary 
solution of eqs.\ (\ref{r5}) and (\ref{r6}) to plot the tunneling 
rate $\tau^{-1}$ numerically. To simplify the discussion, we assume 
that the relaxation rates in the two wells are the same, $\Gamma_1 
=\Gamma_2 \equiv \Gamma$. Figure 4 shows the dependence of 
$\tau^{-1}$ on detuning $\nu$ obtained in this way for several rf 
amplitudes $a$ in the regime of small relaxation rate $\Gamma$. 
The main qualitative feature of Fig.\ 4 (that can also be seen  
in eq.\ (\ref{r7})) is that for small $a$ the resonant peak in 
the tunneling rate $\tau$ is split into two peaks due to coherent 
oscillations of flux between the two wells. The appearance of such 
splitting can be easily understood, since a weak rf signal excites 
the system not into the state $|1\rangle$ localized in the left 
well, but into the two hybridized states formed out of the states 
$|1\rangle$ and $|2\rangle$ in the two wells. The two different 
energies of the two hybridized states lead to the two 
peaks in the tunneling rate. This means that the splitting of the 
resonant-tunneling peak is the direct manifestation of the 
quantum coherent oscillations of flux between the two wells of 
the SQUID potential.

Figure 4 illustrates also how the splitting of the resonant 
tunneling peak is suppressed with the increase of the rf 
amplitude. Suppression of the peak splitting can be understood 
in terms of the time $1/\Delta$ required to establish the 
stationary states hybridized between the two wells. Large 
rf amplitude $a$ on the order of the tunnel amplitude 
$\Delta$ causes rapid Rabi oscillation between the states 
$|0\rangle$ and $|1\rangle$ and does not allow for 
sufficient time to establish the two hybridized states. 
The system is therefore effectively excited into the state 
$|1\rangle$ localized in the left well, and single resonant 
tunneling peak is formed around the energy of this state. 
Broadening and suppression on this single peak seen in 
Fig.\ 4 as $a$ increases beyond $\Delta$, is a version 
of the generic ``quantum Zeno'' effect, when tunneling out 
of a metastable state is suppressed by rapid perturbation 
of this state.

\begin{figure}

\setlength{\unitlength}{1.0in}
\begin{picture}(3.0,2.3) 
\put(0.,0.){\epsfxsize=3.0in\epsfysize=2.3in\epsfbox{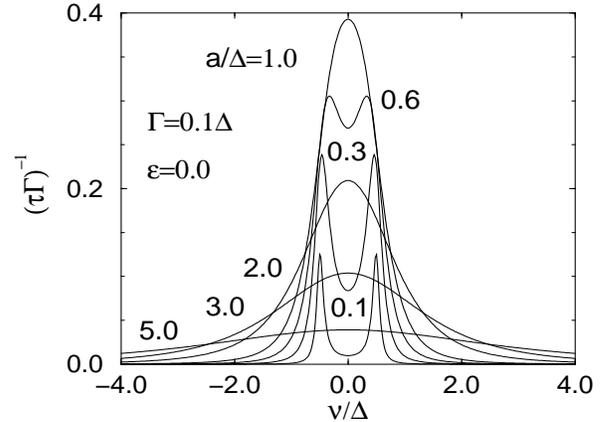}}
\end{picture}

\caption{The rate $\tau^{-1}$ of the photon-assisted resonant 
tunneling of flux $\Phi$ between the two wells of the SQUID 
potential as a function of detuning $\nu$ for several values 
of the rf amplitude $a$. For discussion see the text.}  
\label{fig:4} \end{figure}

Figure 5 shows the evolution of the coherently-split tunneling 
peaks at small rf amplitude $a$ with the bias energy $\varepsilon$ 
and with the relaxation rate $\Gamma$. We see that with increasing 
energy bias (Fig. 5a), the peaks follow the position of the energy 
levels and 
the splitting between them increases. Simultaneously, the peak 
height decreases reflecting the overall suppression of the tunneling 
rate as one moves away from the resonance. Figure 5b shows how the 
double-peak structure in the small-bias regime representing the 
coherent mixing of the flux states in the two wells is suppressed 
by increasing relaxation rate $\Gamma$. The structure is visible 
up to the relaxation rates $\Gamma \simeq 0.5\Delta$. 

The peaks in Fig.\ 5 are shown only for $\varepsilon \geq 0$. The 
peak structure for negative $\varepsilon$ can be understood from the 
``symmetry'' relation $\tau^{-1} (-\nu,-\varepsilon)= \tau^{-1} 
(\nu, \varepsilon)$ that can be deduced from eqs.\ (\ref{r5}) and 
(\ref{r6}). Equations (\ref{r5}) show that in the stationary regime, 
changing the sign of $\nu, \varepsilon$ is equivalent to changing 
the sign  
of $a, \Delta$ and replacing the off-diagonal elements of $\rho$ with 
their complex conjugate values. This transformation obviously does 
not change the transition rates in eq.\ (\ref{r6}), and therefore 
does not change the flux tunneling rate $\tau^{-1}$.  
 
Another interesting manifestation of the coherent flux tunneling 
between the wells can be seen in the dependence of the tunneling 
rate on the bias energy $\varepsilon$ at fixed detuning $\nu$ 
(Fig.\ 6). For weak relaxation, the hybridized states are 
well-developed, and when the rf excitation energy lies between 
these two states, the tunneling rate is strongly suppressed. At 
$\nu=0$, when the system is excited precisely into the state 
$|1\rangle$ localized in the left well, this condition is 
satisfied and tunneling rate is strongly suppressed for any energy 
bias $\varepsilon$. As can be seen from eq.\ (\ref{r7}) and Fig.\ 
6, in this case the tunneling rate as a function of $\varepsilon$ 
is described by a Lorentzian centered around $\varepsilon=0$. 
In contrast to resonant tunneling peaks in the $\nu$-dependence 
of the tunneling rate, which have small width proportional to the 
relaxation rate $\Gamma$,  the width of this Lorentzian is large, 
$\Delta^2/\Gamma$, and is inversely proportional to $\Gamma$. 
When the detuning $\nu$ deviates from zero, there is an energy 
bias $\varepsilon$ at which the excitation energy coincides with 
the energy of one of the hybridized states. The tunneling rate 
has a peak under such resonance conditions. For not-too-small 
$\nu$'s, this resonant peak again has a small width proportional 
to $\Gamma$. In Fig.\ 6, one can see how the transition between 
the broad and narrow tunneling peaks takes place for $\nu \geq 0$. 
As before, the results for negative detuning can be deduced from 
the relation $\tau^{-1}(-\nu,-\varepsilon)= \tau^{-1} (\nu, 
\varepsilon)$. 

\begin{figure}

\setlength{\unitlength}{1.0in}
\begin{picture}(3.0,2.3) 
\put(0.,0.){\epsfxsize=3.0in\epsfysize=2.3in\epsfbox{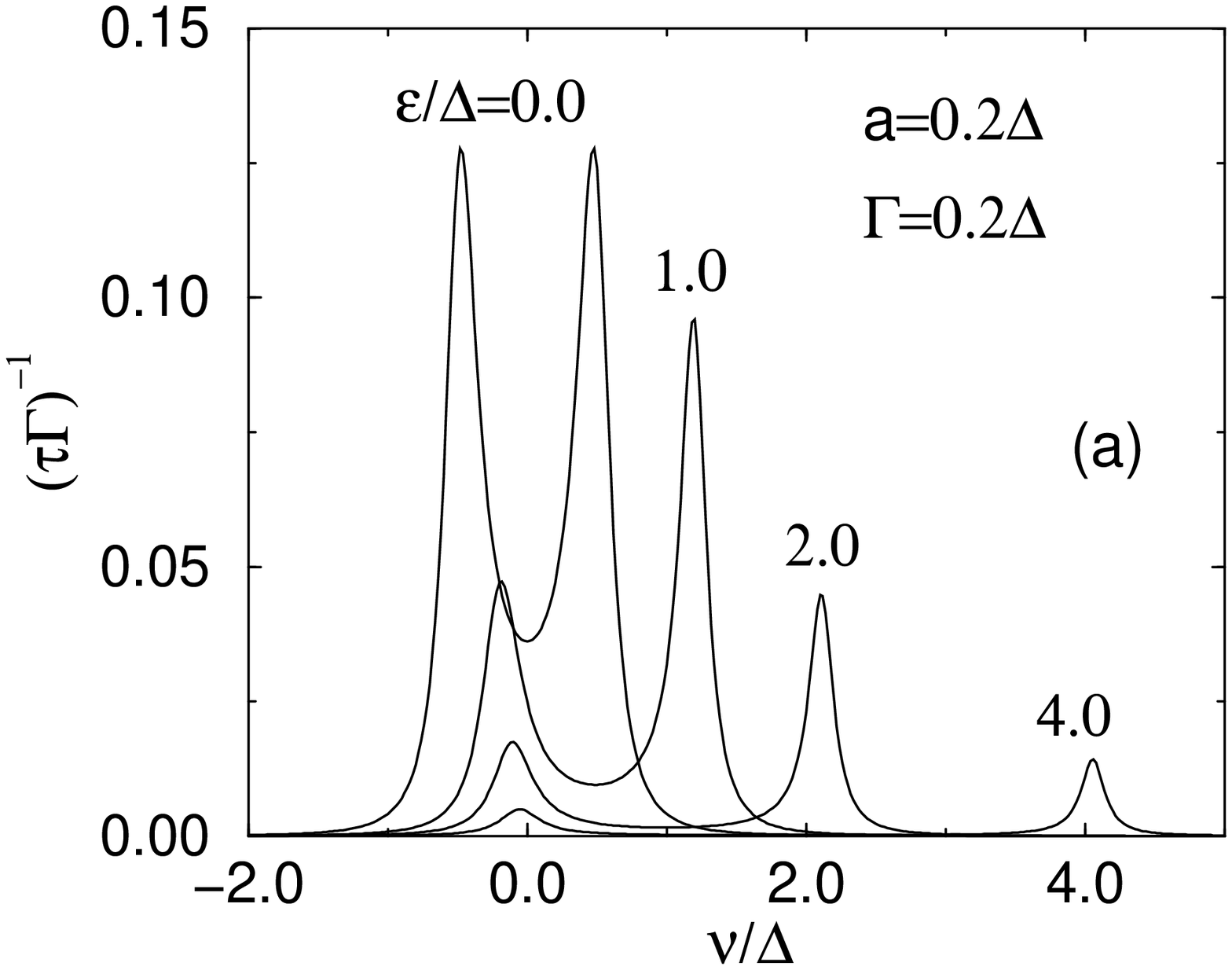}}
\end{picture}

\begin{picture}(3.0,2.3) 
\put(0.,0.){\epsfxsize=3.0in\epsfysize=2.3in\epsfbox{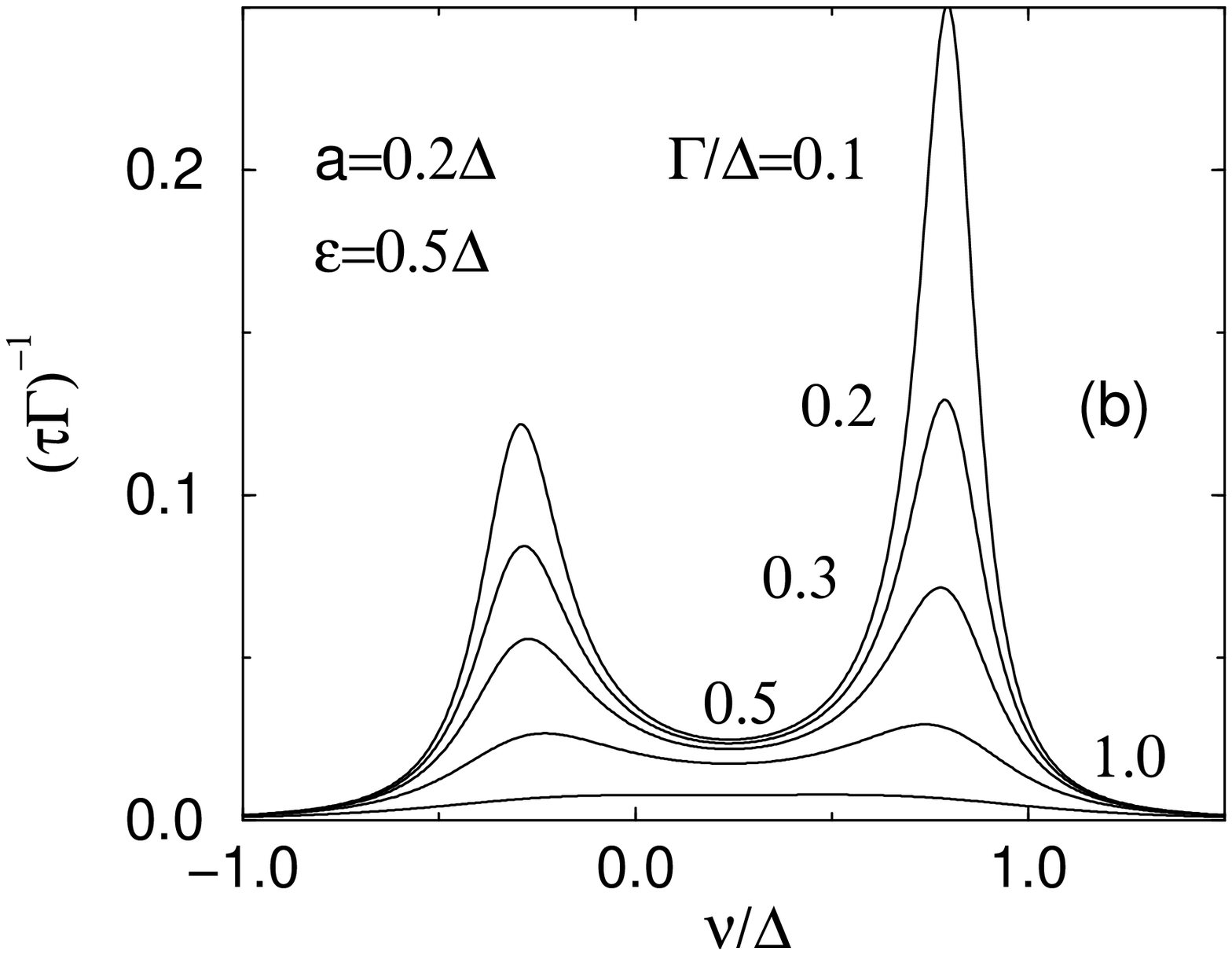}}
\end{picture}

\caption{Evolution of the double-peak structure in the rate 
$\tau^{-1}$ of the photon-assisted resonant flux tunneling as a 
function of detuning $\nu$ with increasing (a) bias energy 
$\varepsilon$, and (b) relaxation rate $\Gamma$. }  
\label{fig:5} \end{figure}

Finally, we discuss the effect of weak interwell relaxation 
on the photon-assisted tunneling. We start by generalizing 
eq.\ (\ref{9}) for this relaxation to the three-state situation 
relevant for the photon-assisted tunneling. Under the natural 
assumption that the average flux in the states $|0\rangle$ and 
$|1\rangle$ in the left well of the SQUID potential is the same, 
the part of the dissipative coupling (\ref{3}) that corresponds 
to the interwell relaxation is: 
\begin{equation}
V=-I_f \delta \Phi \left( \begin{array}{ccc} 1 & 0 & 0 \\ 
0 & 1 & 0 \\ 0 & 0 & -1 \end{array} \right) \equiv -I_f \delta 
\Phi U\, .  
\label{25} \end{equation} 
While eq.\ (\ref{25}) is written in the flux basis 
$|0\rangle\, , |1\rangle\, ,|2\rangle$, weak relaxation is 
conveniently described in the basis of the eigenstates 
$|n\rangle$ of the Hamiltonian (\ref{r2}). In this basis, the 
contribution of the interwell relaxation (\ref{25}) to the 
evolution of the density matrix $\rho$ is given by the 
standard expression similar to eq.\ (\ref{9}): 
\begin{eqnarray}
\dot{\rho}_{nn} & = &  \sum_m ( \gamma_{mn} \rho_{mm} - 
\gamma_{nm} \rho_{nn}) \, ,\label{26}  \\  
\dot{\rho}_{nm} & = &  -[\gamma'_{mn} + \frac{1}{2} 
\sum_k ( \gamma_{nk} + \gamma_{mk}) ] \rho_{nm} \, , \;\;\; 
n\neq m \, . \nonumber  
\end{eqnarray} 
Transition and dephasing rates in these equations are: 
\[ \gamma_{nm} =  \frac{g |U_{nm}|^2 (\varepsilon_n- 
\varepsilon_m)}{1- e^{-(\varepsilon_n-\varepsilon_m) /T}} \, ,  
\;\;\;\; \gamma'_{nm} =  \frac{gT  }{2} 
(U_{nn}-U_{mm})^2\, , \]
where $U_{nm}$ are the matrix elements of the operator $U$ 
(\ref{25}) in the eigenstates basis, and $\varepsilon_n$ is the 
energy of the eigenstate $|n\rangle$.

\begin{figure}

\setlength{\unitlength}{1.0in}
\begin{picture}(3.0,2.3) 
\put(0.,0.){\epsfxsize=3.0in\epsfysize=2.3in\epsfbox{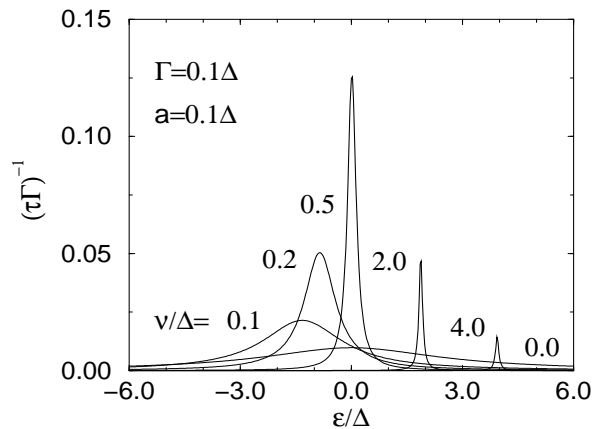}}
\end{picture}

\caption{Bias-energy dependence of the photon-assisted flux 
tunneling rate at fixed detuning $\nu$ for small rf amplitude and 
relaxation rate $\Gamma$. For vanishing detuning, the tunneling 
rate exhibits a very broad maximum with a width inversely 
proportional to $\Gamma$. }  
\label{fig:6} \end{figure}

Interwell relaxation can be included into the evolution equations 
for the density matrix on the basis of eq.\ (\ref{26}) numerically. 
We diagonalize the Hamiltonian (\ref{r2}),  
calculate the interwell relaxation terms (\ref{26}) in the  
eigenstates basis, and transfer them into the flux basis, where 
the intrawell relaxation has the simple form (\ref{r5}), (\ref{r6}). 
Calculating finally the stationary value of the 
density matrix $\rho$ we find the flux tunneling rate (\ref{tau}). 

Figures 7 and 8 show results of such a calculation obtained at 
vanishing temperature $T$. In Figure 7, the tunneling rate is 
plotted as a function of the detuning $\nu$ 
for $\varepsilon=0$ and several values of the relative strength of 
the intrawell relaxation $\Gamma$. The eigenstates of the 
two-state Hamiltonian at $\varepsilon=0$ are symmetric between the 
two wells, and in absence of the interwell relaxation produce two 
symmetric resonant tunneling peaks (see Figs.\ 4 and 5a). As can be 
seen from Fig.\ 7, the interwell relaxation makes the tunneling 
peaks asymmetric. The positive-$\nu$ side of the double-peak 
structure corresponds to excitation of the system into the 
lower-energy eigenstate and is unaffected by the interwell 
relaxation at zero temperature, since there is no energy in this 
regime to create additional tunneling path. In contrast, the 
negative-$\nu$ side of the double-peak structure corresponds to 
excitation of the system into the eigenstate with larger energy, 
and the interwell relaxation increases the rate of tunneling out 
of this state. Because of this, the negative-$\nu$ peak in Fig.\ 7 
is larger than the tunneling peak at positive $\nu$, and the 
tunneling rate at $\nu<0$ decreases much more slowly away from 
the peak than at positive $\nu$. 

\begin{figure}[htb]
\setlength{\unitlength}{1.0in}
\begin{picture}(3.0,2.3) 
\put(0.,0.){\epsfxsize=3.0in\epsfysize=2.3in\epsfbox{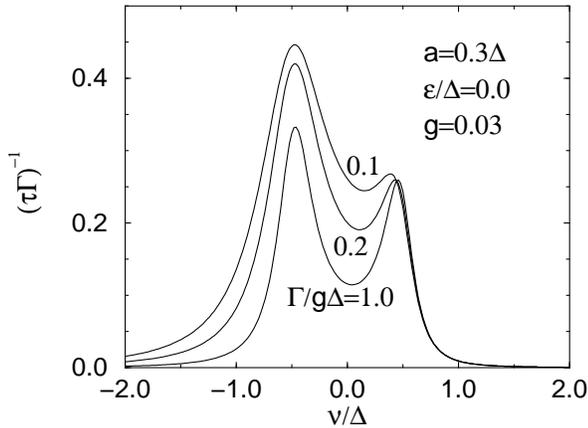}}
\end{picture}
\caption{The rate $\tau^{-1}$ of the photon-assisted flux tunneling 
as a function of the detuning $\nu$ in the case of symmetric 
coupling between the tunneling flux states, $\varepsilon=0$, in the 
presence of both interwell and intrawell relaxation. Different 
curves correspond to different magnitudes of the intrawell relaxation 
rate $\Gamma$ relative to the interwell relaxation rate. }
\end{figure} 

Interwell relaxation introduces asymmetry also in the dependence 
of the flux tunneling rate on the bias energy $\varepsilon$. 
Examples of such dependence are shown in Fig.\ 8 for two values of 
the detuning, $\nu=\pm 2\Delta$. In both cases, the tunneling rate 
has a resonant peak at $\varepsilon \simeq \nu$ similar to the 
peaks shown in Fig.\ 6 for vanishing interwell relaxation, when 
the peaks for the two values of detuning are symmetric. Comparison 
of Figs.\ 8a and 8b shows that the interwell relaxation makes the 
peaks asymmetric. In particular, the peak at $\nu<0$ is smaller 
than the peak at $\nu>0$. Although this asymmetry appears to 
be opposite to that in Fig.\ 7, where the negative-$\nu$ peak is 
larger, it has the same origin as in Fig.\ 7. The peaks at $\nu<0$ 
and $\nu>0$ correspond to resonant excitation of the system into, 
respectively, the upper and lower energy eigenstates. When the 
resonance occurs for $|\varepsilon|>\Delta$, as in Fig.\ 8, the 
eigenstates are already to a large extent localized in one or the 
other well. At $\varepsilon \simeq \nu >0$, the lower eigenstate is 
centered in the right well and the interwell relaxation increases 
the tunneling rate, while at $\varepsilon \simeq \nu <0$ the lower 
eigenstate is centered in the left well and the interwell 
relaxation brings the system back to this well suppressing the 
tunneling rate. As a result, the resonant tunneling peak in Fig.\ 
8a ($\nu >0$) is larger than in Fig.\ 8b ($\nu <0$). The height 
of the negative-$\nu$ peak is more sensitive to the relative 
strength of the two relaxation mechanisms and decreases with 
decreasing rate $\Gamma$ of the intrawell relaxation.  

\begin{figure}

\setlength{\unitlength}{1.0in}
\begin{picture}(3.0,2.3) 
\put(0.,0.){\epsfxsize=3.0in\epsfysize=2.3in\epsfbox{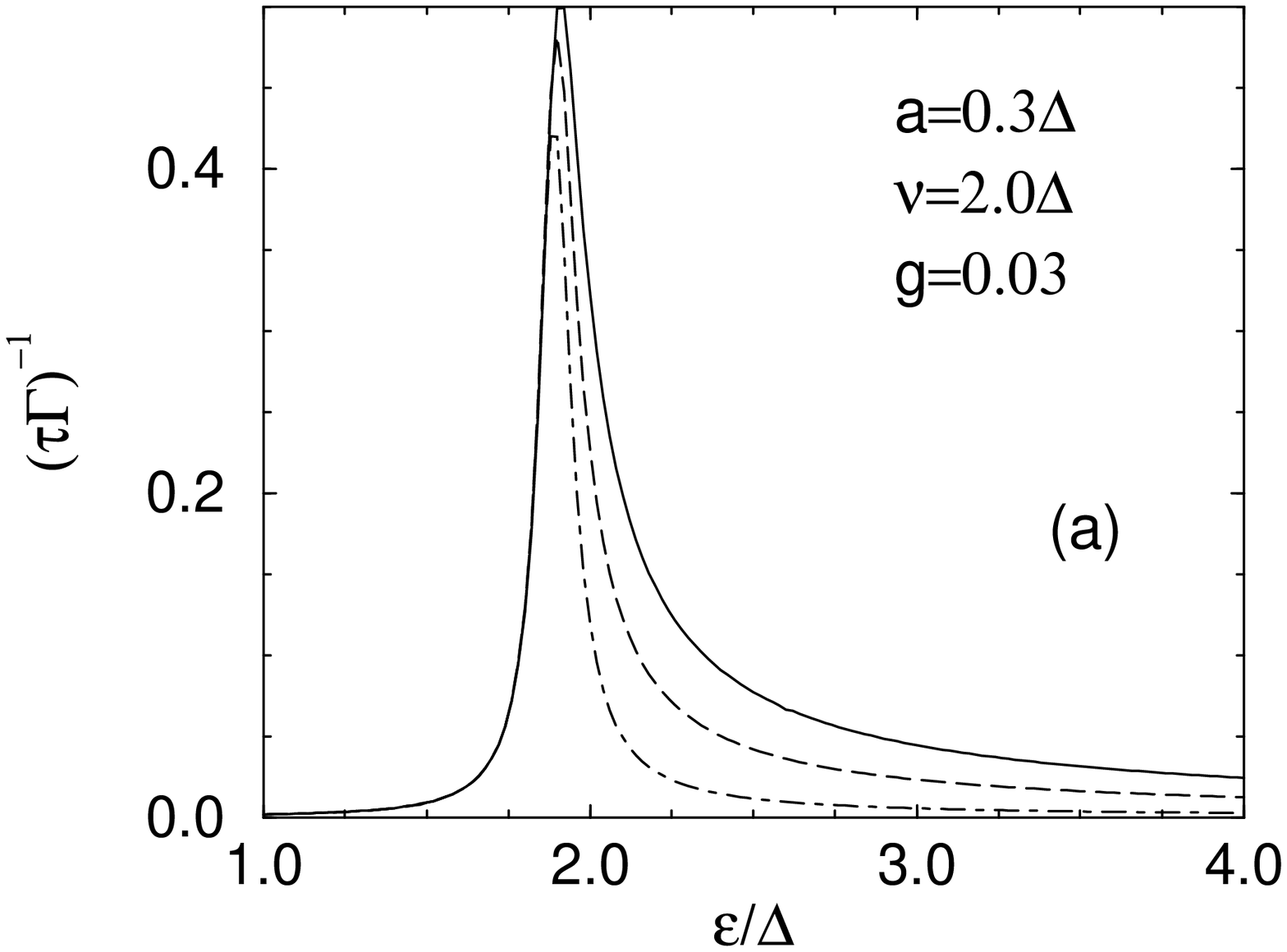}}
\end{picture}

\begin{picture}(3.0,2.3) 
\put(0.,0.){\epsfxsize=3.0in\epsfysize=2.3in\epsfbox{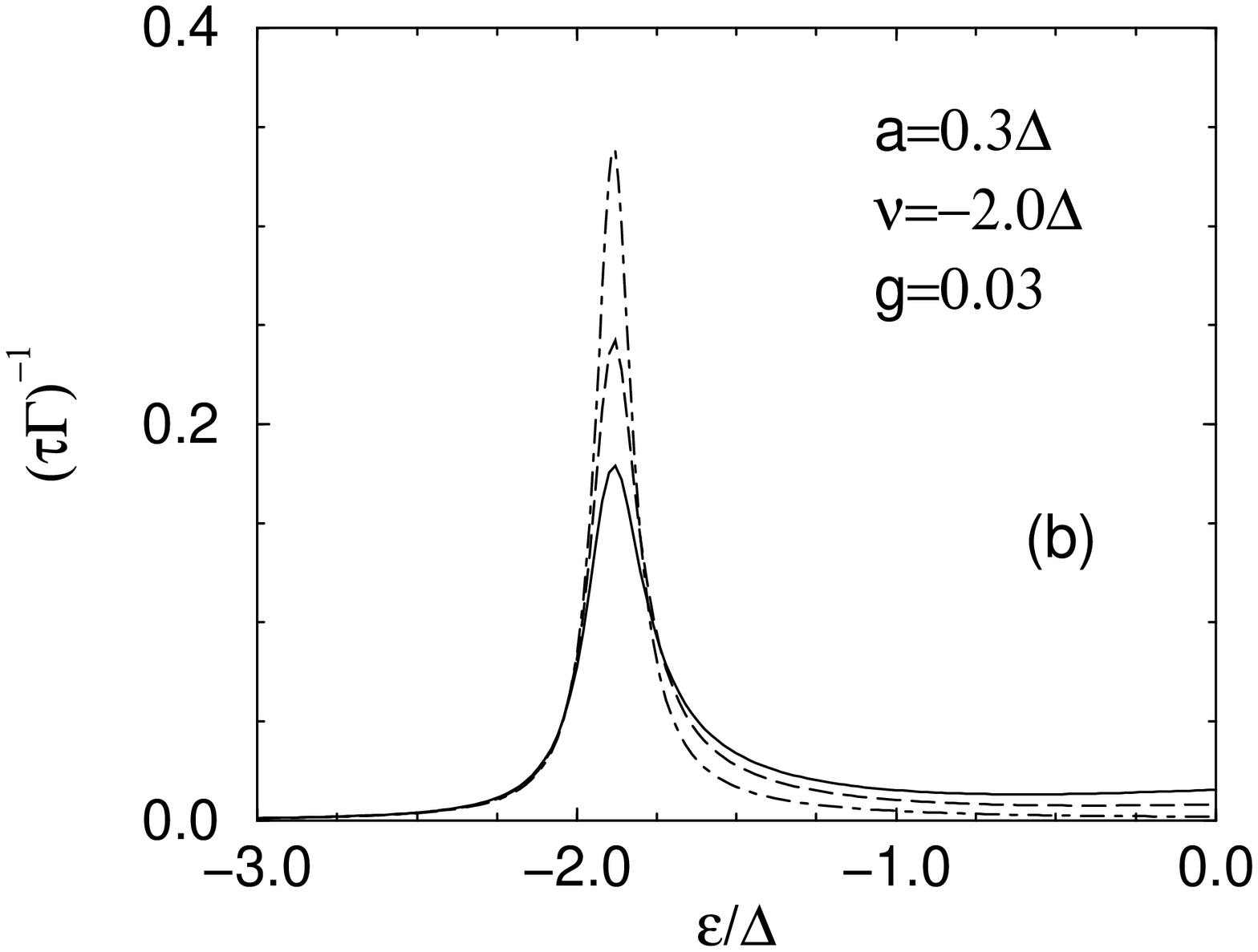}}
\end{picture}

\caption{The rate $\tau^{-1}$ of the photon-assisted flux tunneling 
as a function of the bias energy $\varepsilon$ for two ``symmetric'' 
values of the detuning $\nu$ and the same rates of the intrawell 
relaxation $\Gamma$ as in Fig.\ 7. In (a), $\tau^{-1}$ decreases 
with increasing $\Gamma$ for $\varepsilon$ above the peak, while in (b), 
the peak height increases with $\Gamma$. }  
\label{fig:8} \end{figure}

The tails of the photon-assisted resonant peaks can also be 
described analytically. When the bias energy and detuning are not 
close to any resonance, $|\nu|,|\varepsilon|,|\nu-\varepsilon|\gg 
a,\Delta$ both the interwell tunneling and rf excitation can be 
treated as perturbations. The dynamics of flux tunneling in this 
regime can be described as a coexistence of the two tunneling 
paths similar to the off-resonant tunneling discussed in Sec.\ 2 
(see inset to Fig.\ 2). If $\nu >\varepsilon $, the effective 
energy $\nu-\varepsilon$ of the state $|2\rangle$ in the right 
well (i.e., the energy of this state brought down by a quantum 
of rf radiation, as in the Hamiltonian (\ref{r2})) is above the 
energy of 
the initial state $|0\rangle$ in the left well, and the only 
energy-allowed tunneling path is direct relaxation in the right 
well out of the perturbed state $|0\rangle$. Similarly to eq.\ 
(\ref{16a}), perturbation theory in $a,\,\Delta$ gives for the 
rate $\bar{\Gamma}$ of this tunneling: 
\begin{equation} 
\bar{\Gamma}=\frac{\Gamma_2 a^2 \Delta^2 }{16\nu^2(\nu- 
\varepsilon)^2} \, . 
\label{30} \end{equation} 
If $\nu <\varepsilon$, the effective energy of the state 
$|2\rangle$ is lower than the energy of the state $|0\rangle$, 
and in addition to the tunneling (\ref{30}) there is a competing 
tunneling process. It consists of a transition between the perturbed 
states $|0\rangle$ and $|2\rangle$ with the rate $\bar{\gamma}$,  
\begin{equation} 
\bar{\gamma}= \frac{g a^2\Delta^2 }{4\nu^2 (\varepsilon-\nu) } \, .
\label{31} \end{equation} 
that is driven by interwell relaxation, followed by direct 
relaxation in the right well with the rate $\Gamma_2$. As in Sec.\ 
2, the coexistence of the two tunneling paths gives the following 
total tunneling rate: 
\begin{equation} 
\tau^{-1}= \Gamma_2 \frac{\bar{\gamma}+ \bar{\Gamma}}{\bar{\gamma} 
+\Gamma_2} \, .
\label{32} \end{equation}
One can check that eqs.\ (\ref{30}) and (\ref{32}) agree, 
respectively, with the negative-$\epsilon$ tail of the resonant 
tunneling peak in Fig.\ 8b and the positive-$\epsilon$  
tail of the peak in Fig.\ 8a. 

In summary, we have studied the effects of two types of relaxation 
mechanisms on the macroscopic resonant tunneling of flux in SQUIDs 
under stationary-bias conditions and with external rf 
irradiation. Coherent splitting of the resonant-tunneling peaks 
by rf radiation provides a convenient way of studying quantum 
coherence of flux states.

\vspace{1ex} 

This work was supported by ARO grant DAAD199910341. 

\vspace{3ex}   
 
{\bf Appendix} 

\vspace{1ex} 

In the Appendix we show explicitly how the Hamiltonian of the 
two-well system (\ref{1}) can be reduced at resonance to the 
two-state form (\ref{2}), and derive an expression for the 
tunneling amplitude $\Delta$. Assuming that the transparency $D$ 
of the barrier separating the two wells is small, $D\ll 1$, and 
that the resonance occurs between the states with large $n$, we 
can use the WKB approximation for the wavefunctions $\psi_{jn} 
(\Phi)$ of the Hamiltonian (\ref{1}). In this approximation, 
the wavefunction between the right and the left turning points 
$r$ and $l$ is: 
\begin{equation} 
\psi(\Phi)= \frac{A}{\sqrt{p} }\cos (w(\Phi) -\delta) \, , 
\label{a1} \end{equation} 
with the WKB phase $w(\Phi) =(1/\hbar) \int_{l}^{\Phi} d\Phi'p(\Phi') 
-\pi/4$ and momentum $p=[2C(E-U(\Phi))]^{1/2}$. In eq.\ (\ref{a1}), 
$\delta$ is a constant phase shift, and $A$ is a normalization 
constant. The phase $w$ is defined in such a way that $\delta=0$ 
for an isolated well, when the wavefunction decays exponentially 
in the classically inaccessible region $\Phi>r, \; \Phi<l$. When 
the two wells are coupled, the energy $E$ of the state common to 
them deviates from the eigenenergies $\varepsilon_n$ of the 
isolated wells that are determined by the Bohr-Sommerfeld 
condition $(1/\hbar) \int_{l}^{r} p d\Phi =2\pi (n+1/2)$. For 
weak tunneling, this deviation is small in comparison to 
the state energy, and creates a small but non-vanishing phase 
shift $\delta$: 
\begin{equation}
\delta= \frac{1}{\hbar} \int_l^r (p(E)-p(\varepsilon_n)) = 
\pi \frac{E-\varepsilon_n}{\omega_p} \, ,  
\label{a2} \end{equation} 
where $\omega_p$ is the frequency of the classical oscillations in 
the well and which in the WKB approximation determines the spacing 
of the energy levels. 

At non-vanishing $\delta$, the wavefunction (\ref{a1}) has a part 
that grows exponentially in the classically inaccessible region, 
as one can see rewriting it as 
\begin{equation} 
\psi= \frac{Ae^{i\delta} }{\sqrt{p} }\cos w(\Phi) 
-\frac{i A\sin \delta }{\sqrt{p} } e^{iw(\Phi)} \, . 
\label{a3} \end{equation} 
According to general rules of the WKB approximation -- see, e.g.,
\cite{ll}, the two terms in this expression produce exponentially 
decaying and growing components of the wavefunctions with the 
amplitudes $Ae^{i\delta}/2$ and $-A\sin \delta$, respectively.
To find the energy $E$ we need to match the amplitudes of the 
wavefunctions of the right 
and left wells in the barrier region. Under the conditions of 
resonance, equating the amplitudes and keeping only the terms of 
the first order in $D$ we get: 
\begin{equation} 
A_1\frac{E-\varepsilon_1}{\omega_1} =-A_2\frac{D}{2\pi}\, , 
\;\;\;\;\; 
A_2\frac{E-\varepsilon_2}{\omega_2} =-A_1\frac{D}{2\pi}\, ,
\label{a4} \end{equation} 
where $\varepsilon_j$, $j=1,2$ are the energies of the resonant 
states, and $D=\exp \{-(1/\hbar) \int_{r_1}^{l_2} |p| d\Phi \}$. 

The probability to be in the right/left well is directly 
related to the wavefunction amplitudes $A_j$: 
\begin{equation}  
\int_{l_j}^{r_j} d\Phi |\psi_j|^2 = \frac{B|A_j|^2}{\omega_j} \, , 
\label{a5} \end{equation}  
where $B$ is the $j$-independent part of the normalization 
constant. Introducing the amplitudes $\alpha$ of this 
probability, $\alpha_j=A_j/\omega_j^{1/2}$, we bring eq.\ 
(\ref{a4}) into the form that coincides with the Schr\"{o}dinger 
equation of the two-state system: 
\begin{equation} 
E\alpha_1 =\varepsilon_1 \alpha_1 -\frac{\Delta}{2} \alpha_2 \, , 
\;\;\;\;\; 
E\alpha_2 =\varepsilon_2 \alpha_2 -\frac{\Delta}{2} \alpha_1 \, ,
\label{a6} \end{equation} 
with the tunneling amplitude 
\[ \Delta= \frac{(\omega_1 \omega_2)^{1/2} }{\pi} D \, . \] 
Since the resonant states $|1,2\rangle$ are orthogonal to all 
other states of the full Hamiltonian (\ref{1}), this proves that 
their dynamics can be described by the Hamiltonian (\ref{2}).   

Away from resonance, when $\varepsilon \equiv \varepsilon_1- 
\varepsilon_2 \sim \omega_p$, one of the amplitudes $\alpha$ 
(for the states localized in the left well, $\alpha_2$) is small, 
$\alpha_2 \sim D$. Keeping, as before, only the terms of the first 
order in $D$, we see from eq.\ (\ref{a6}) that in this case 
$E_1= \varepsilon_1$, i.e. $\delta_1=0$. Matching the growing 
wavefunction in the right well with the decaying wavefunction 
in the left well we find that away from the resonance 
\begin{equation}   
\alpha_2 = \frac{\pi \Delta}{ 2\omega_2 \sin (\pi 
\varepsilon/\omega_2) } \, . 
 \label{a7} \end{equation}  
Equation (\ref{a7}) extrapolates smoothly between the successive 
resonances. 

Making use of the harmonic approximation for the potential 
$U(\Phi)$, one can show that the results of this Appendix (eqs. 
(\ref{a6}) and (\ref{a7})) can be extended to the low-lying 
states with small $n$ which cannot be described with the WKB 
approximation. Such an extension leads only to minor 
modifications in the definition of the barrier transparency 
$D$ in these equations.

\newpage

\end{document}